# Stock Price Prediction and Traditional Models: An Approach to Achieve Short-, Medium- and Long-Term Goals

**Opeyemi Sheu Alamu[1*], Md Kamrul Siam[2*]**

[1]Department of Statistics, Federal College of Animals Health and Production Technology, Ibadan, Nigeria
[2]New York Institute of Technology, New York, USA
Email: Sheuopeyemi99@gmail.com, Ksiam01@nyit.edu

## Abstract

A comparative analysis of deep learning models and traditional statistical methods for stock price prediction uses data from the Nigerian stock exchange. Historical data, including daily prices and trading volumes, are employed to implement models such as Long Short Term Memory (LSTM) networks, Gated Recurrent Units (GRUs), Autoregressive Integrated Moving Average (ARIMA), and Autoregressive Moving Average (ARMA). These models are assessed over three-time horizons: short-term (1 year), medium-term (2.5 years), and long-term (5 years), with performance measured by Mean Squared Error (MSE) and Mean Absolute Error (MAE). The stability of the time series is tested using the Augmented Dickey-Fuller (ADF) test. Results reveal that deep learning models, particularly LSTM, outperform traditional methods by capturing complex, nonlinear patterns in the data, resulting in more accurate predictions. However, these models require greater computational resources and offer less interpretability than traditional approaches. The findings highlight the potential of deep learning for improving financial forecasting and investment strategies. Future research could incorporate external factors such as social media sentiment and economic indicators, refine model architectures, and explore real-time applications to enhance prediction accuracy and scalability.

*These authors contributed equally to this work.

## 1. Introduction

The nature of prediction in the ever-changing environment of finance has for long been a core aim in the quest for considering goals for the investor, the financial analyst or even the decision maker. The more conventional methods of analyzing the stock include the statistical and econometric models, which have provided most of the direction and forecast for stock prices, besides helping in the understanding of the existing market trends. However, the development and complexity of today's financial markets have rather imposed certain restrictions on the application of these models and stimulated the increase of efforts toward finding more accurate tools for various investment management approaches.

The pragmatic research question of this work is to discover whether traditional modeling techniques combined with modern machine learning and deep learning approaches can be put into practice in the area of stock price prediction. To this end, the research will seek to analyze their performance on a short, medium and long-term basis with the aim of establishing their overall characteristics and appropriateness for use in various investment goals.

Therefore, the justification for undertaking this research rests in appreciation of the expanding role of knowledge in the management of risks in the financial markets. It is for this reason that giving accurate predictions on the stock price can help the investors to make the correct investment choices, manage the portfolio and reduce risk. Furthermore, employing such knowledge, the financial institutions and the policy makers can design more suitable trading plans and enact rational policies which support the stability and the further evolution of such markets.

Previous research has discussed the use of such approaches as time-series analysis, regression models, and learning machines for the construction of stock price forecasts. Although these studies have been beneficial in nature, there is still a need to perform a comparative study of the returns from the traditional model, as well as the new-applied techniques in the short-term, medium term, and the long-term investment horizon [1]-[4].

This present research seeks to mitigate this fact by undertaking a thorough evaluation of the conventional models like ARIMA, GARCH, and Vector Autoregressive (VAR) models alongside new-class machine learning algorithms like Artificial Neural Networks (ANNs), Support Vector Machines, and Long Short-Term Memory (LSTMs). Hence, the research will compare and understand and compare the predictive capabilities and consistency of these approaches across multiple horizons thereby providing the foundation on which investors and financial experts can select the best approach for the achievement of the intended goals in investment.

This work will examine whether it is possible to improve stock price prediction by using some mid-level models and integrate them or use deep learning methods on the same models. The outcomes of this study will expand the need-to-know base of financial forecasting scholarship and supply usable guidance for the investor, financial analysts, and policymakers who want to better understand the behavior of the stock market.

Therefore, this study provides a detailed analysis of the efficacy in traditional model and new methods for stock price prediction with special consideration for meeting the demand of short-term, medium-term, and long-term investment horizons. This work will be useful to practitioners as well as researchers since the findings will be an invaluable resource in the anxiety-inspiring world of financial markets, and the decisions made within them.

## 2. Literature Review

### 2.1. Introduction

This literature review explores the existing research on stock price prediction, focusing on the use of both deep learning and traditional models. By reviewing the current state of the art, this section aims to identify the strengths and limitations of each approach and highlight areas for further improvement. Understanding the theoretical foundations and empirical findings of these models is essential for developing a comprehensive framework for stock price prediction that leverages the strengths of both deep learning and traditional techniques.

### 2.2. Overview of Stock Price Prediction with Models

The interest in stock market analysis is not limited to economics; it also attracts researchers from diverse backgrounds, including engineering disciplines. While most studies focus on stock market analysis, recent works have expanded to predicting forex and cryptocurrencies, reflecting the dynamic nature of financial markets [5] [6]. Cryptocurrencies, in particular, present unique challenges due to their decentralized structure and lack of regulatory oversight, making them more susceptible to manipulation and fraud compared to traditional stock markets.

Research has incorporated machine learning and more precisely deep learning for predicting the stock price and cryptocurrency. Previous research established that RNN effectively predicted daily Bitcoin prices better than other approaches such as ARIMA because of its capability to consider time-related characteristics of the data [7]. The studies conducted by the researchers have shown that although ARIMA generally presents high accuracy it needs stationary data as opposed to non-stationary time series handled by machine learning algorithms.

Also, a short-term stock prices forecasting model based was applied upon ARIMA, on three sectors based on the National Stock Exchange [8]. From their findings, the two identified possible gains in floating shares in the FMCG and Pharmaceuticals industries but possible loss in the Banking industry. In another study using data from Dhaka Stock Exchange was explored where they used a new univariate ANN model that overpowers the conventional ARIMA model for forecasting the total market capitalization analogously [9].

Further contributing to the area of work, some work was done on the impact that hybrid models and gated recurrent units (GRUs) have in enhancing stock price prediction [10] [11]. Those approaches were extended the study by using a GRU-based model coupled with a procedure of dataset reconstruction to address overfitting issues [12]. These works demonstrate the effectiveness of implementing deep learning techniques in the stock market prediction despite there are some approaches that highlight that such patterns are not the only factors that must be taken into consideration for accurate prediction of the stock markets [13] [14].

The problem of predicting stock prices has been deemed difficult due to the unstructured and non-stationary behavior of financial markets. Models based on statistical analysis have many drawbacks with regards to identifying the retrospective and probabilistic characteristics of patterns characteristic of stock markets. As for recent trends, authors have tried to use machine learning and deep learning to enhance the efficiency of stock price prediction.

Also, the similar line was employed by Nabipour *et al.* using decision trees, boosting techniques, and LSTM in an effort to classify the stock market group of Tehran Stock Exchange [15]. Their outcome showed that the LSTM model has better performance than the other techniques, which pointed out the use of deep learning in stock market prediction.

More recently, Liu *et al.*, and Nikou *et al.* discovered that stock market indices and exchange-traded fund prices were better predicted by LSTM-based recurrent neural networks than simpler predictive models [2] [4]. These studies highlighted LSTM's performance in modeling long-term dependencies in stock market data.

Moreover, Vijh *et al.* evaluated the factors of artificial neural network (ANN) and random forest (RF) and proved that ANN had better prediction performance than that of RF for some companies, which supported that ANN can be used for stock price prediction [3]. Hu *et al.* surveyed the literature regarding the recent increase in the use of LSTM fused with other deep learning techniques for financial forecasting and found that the use of deep learning methods is increasing continuously [16].

Moghar *et al.* identified the application of LSTM in forecasting future stock market values, and this established the potential of recurrent neural networks in this field [17]. In a nutshell, the studies' findings stress on the benefits of deep learning, especially LSTM, for improving the forecast of stock price and providing requisite information to investors and policymakers. Nti *et al.* conducted a comprehensive review of 122 papers published between 2007 and 2018 on stock market prediction [18]. They found that 66% of the papers used technical analysis methods, while 23% and 11% used fundamental analysis and combined analysis methods, respectively. Additionally, 98% of the fundamental analysis studies used data from social network sites to infer sentiment on financial markets. Commonly used technical indicators included SMA, EMA, MACD, RSI, and ROC.

### 2.3. Comparison between Deep Learning and Traditional Models

The field of stock price prediction has seen significant advancements with the introduction of deep learning models, particularly in comparison to traditional models like autoregressive integrated moving averages (ARIMA) and artificial neural networks (ANN). This comparison has been explored in various studies, shedding light on the strengths and limitations of each approach.

Qihang compares three models: ARIMA, ANN, and LSTM networks, for stock price prediction [19]. The study suggests that LSTM may possess superior predictive abilities, contingent on effective data processing. Additionally, the ANN model outperforms the ARIMA model, highlighting the importance of incorporating time series data with external factors for improved predictions. Despite the advancements in LSTM models, all three models fundamentally derive predictions from potential relationships within time series data, neglecting external factors such as economic and political dynamics. The study emphasizes the need for future research to delve deeper into understanding and incorporating these external factors into predictive models for more accurate forecasting.

Seabe *et al.* focus on predicting cryptocurrency prices using deep learning models, including LSTM, Gated Recurrent Unit (GRU), and Bi-Directional LSTM (Bi-LSTM) [20]. The study concludes that accurate cryptocurrency predictions are crucial, emphasizing the challenge posed by the market's nonlinearity. The evaluation identifies Bi-LSTM as the most effective model, outperforming LSTM and GRU. The study suggests that incorporating external factors such as social media and trading volumes could further enhance prediction accuracy. Additionally, a cost–benefit analysis acknowledges the expenses involved in building and maintaining prediction models but underscores the potential revenue generation and valuable predictions as significant benefits.

Both studies highlight the advancements made by deep learning models, particularly LSTM and Bi-LSTM, in stock price and cryptocurrency price prediction, respectively. These models have shown superior performance compared to traditional models like ARIMA and ANN. However, both studies also emphasize the importance of incorporating external factors into predictive models to improve accuracy further. While deep learning models have shown great promise, there is still ample room for further research and development to enhance their capabilities and address the complexities of financial markets.

## 3. Method

This research focuses on the performance of deep learning models and in particular Recurrent Neural Network (RNNs) including LSTM and GRU as well as benchmarking them against conventional statistical models like ARMA and ARIMA in the context of Nigerian stock market price prediction.

### 3.1. Scope of the Research

The study aims at predicting the daily stock prices over a Five-year period starting from January 1, 2019, to December 31, 2023. The study tries to establish the models' comparative performance in diverse scenarios by assessing their accuracy at various time frames, such as short-term (one year), mid-term (two and a half years), and long-term (five years).

Two goals of the project are to compare the accuracy of deep learning models with standard statistical models in order to produce helpful recommendations for risk managers and investors in Nigeria's stock market. In as much as it seeks to identify the relative performance of these models there is a noble goal of providing direction to this decision making and improving investment strategies, thus the researchers' desire of advancing the use of financial forecasting techniques. Figure 1 provides a clear and detailed illustration of how the deep learning method operates for stock prediction, aiding in a comprehensive understanding of its functionality.

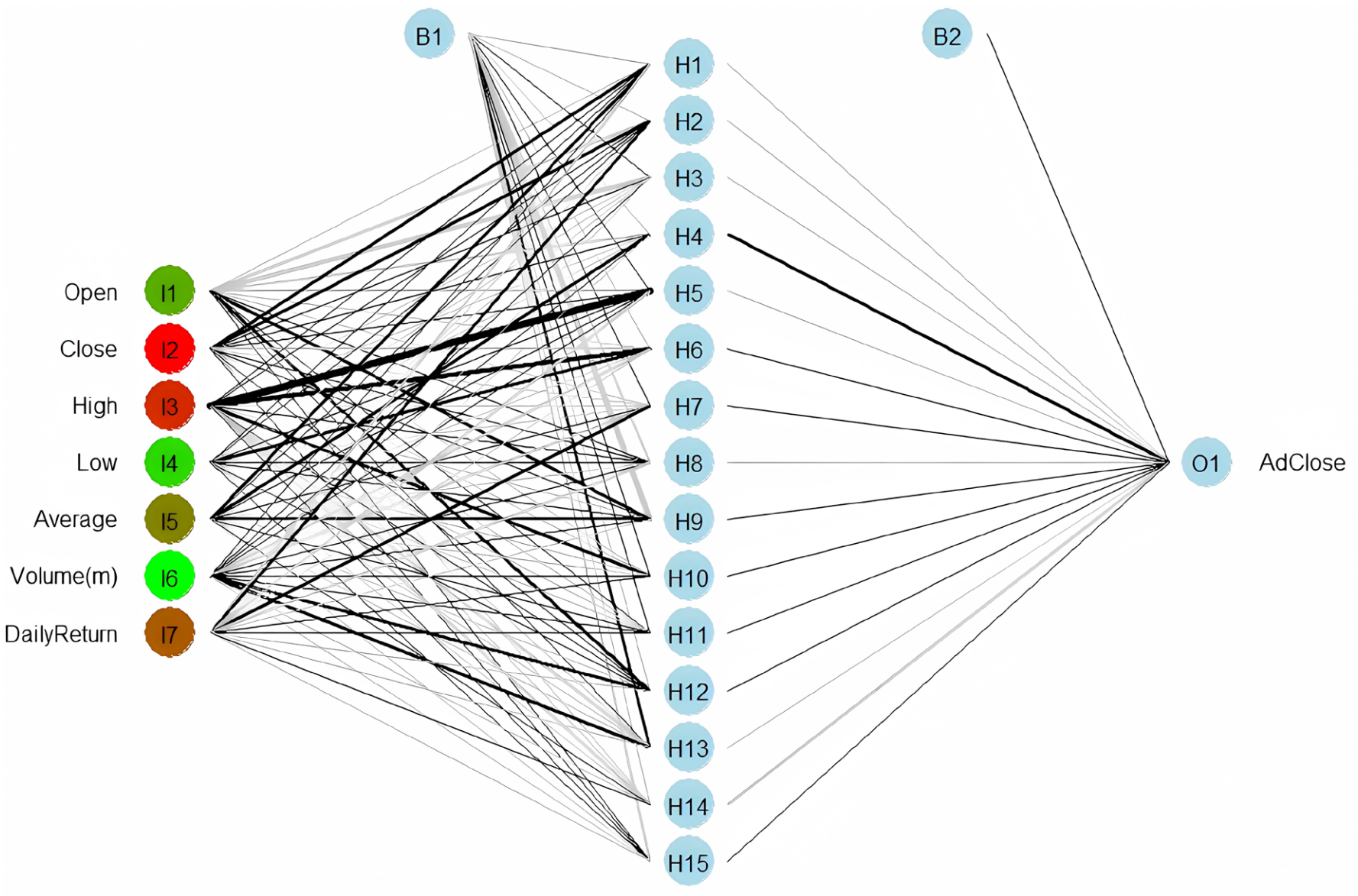

Figure 1. The architecture of deep learning model of stock price.

The use of this methodological framework will lay a good platform for proper assessment of any of the selected models as will be seen in the subsequent analysis and comparison of their forecasting capability and reliability in the Nigerian stock market.

## 3.2. Source and Description of Dataset

The dataset used in this study was obtained from reliable financial databases and covers daily stock prices for the specified period. It includes key parameters such as opening and closing prices, high and low prices, and trading volumes. To ensure the data's accuracy and real-time updates, the study utilized the Yfinance API, which is known for providing comprehensive and dynamic financial data. This API forms a solid foundation for the analysis, crucial for the study's success. The dataset's reliability and comprehensiveness are vital, given its columns for Date, Open, High, Low, Close, Adjusted Close, and Volume. The Date column records the trading day, Open and Close prices reflect the stock's prices at market opening and closing, High and Low indicate the day's highest and lowest prices, Adjusted Close adjusts for corporate actions such as dividends, and Volume denotes the number of shares traded. Table 1 and Table 2 represents the pattern and summary of the data respectively.

Table 1. Dataset retrieved from the Yfinance (From January 2019 to February 2019).

| Date | Open | High | Low | Close | Adj Close | Volume |
|---|---|---|---|---|---|---|
| **2019-01-02 00:00:00** | 16.4 | 16.5 | 16.19 | 16.5 | 7.778549 | 5133 |
| **2019-01-03 00:00:00** | 16.23 | 16.98 | 15.91 | 15.91 | 7.500408 | 8671 |
| **2019-01-04 00:00:00** | 15.64 | 16.98 | 15.64 | 16.493 | 7.775249 | 2837 |
| **2019-01-07 00:00:00** | 15.92 | 16.2 | 15.92 | 16.1 | 7.589979 | 7249 |
| **2019-01-08 00:00:00** | 16.48 | 17.09 | 15.4 | 15.8 | 7.44855 | 23243 |
| **2019-01-09 00:00:00** | 15.92 | 16.5728 | 15.9 | 15.9 | 7.495693 | 9248 |
| **2019-01-10 00:00:00** | 15.97 | 16.2537 | 15.97 | 16.01 | 7.547551 | 6631 |
| **2019-01-11 00:00:00** | 16.02 | 16.3097 | 16.02 | 16.16 | 7.618264 | 1239 |
| **2019-01-14 00:00:00** | 16.25 | 16.37 | 16.2 | 16.2 | 7.637122 | 9561 |
| **2019-01-15 00:00:00** | 16.38 | 16.4 | 16.31 | 16.4 | 7.731407 | 2700 |
| **2019-01-16 00:00:00** | 16.3827 | 16.442 | 16.31 | 16.34 | 7.703121 | 1823 |
| **2019-01-17 00:00:00** | 16.4084 | 16.4899 | 16.4084 | 16.4841 | 7.771054 | 1539 |
| **2019-01-18 00:00:00** | 16.48 | 16.8246 | 16.48 | 16.721 | 7.882737 | 11415 |
| **2019-01-22 00:00:00** | 16.57 | 16.57 | 16.1 | 16.1 | 7.589979 | 8723 |
| **2019-01-23 00:00:00** | 16.22 | 16.72 | 16.13 | 16.544 | 7.799292 | 11842 |
| **2019-01-24 00:00:00** | 16.5 | 16.55 | 16.24 | 16.5202 | 7.788074 | 3602 |
| **2019-01-25 00:00:00** | 16.78 | 17.15 | 16.76 | 17.1 | 8.061406 | 13156 |
| **2019-01-28 00:00:00** | 17.15 | 17.5847 | 17.15 | 17.15 | 8.084977 | 4834 |
| **2019-01-29 00:00:00** | 16.99 | 17.286 | 16.99 | 17.01 | 8.018978 | 18050 |
| **2019-01-30 00:00:00** | 16.94 | 17.2047 | 16.81 | 17 | 8.014263 | 2802 |
| **2019-01-31 00:00:00** | 16.6 | 16.9201 | 16.57 | 16.65 | 7.849263 | 4780 |
| **2019-02-01 00:00:00** | 16.65 | 16.8 | 16.6 | 16.6 | 7.825693 | 3043 |
| **2019-02-04 00:00:00** | 16.85 | 17.0612 | 16.85 | 16.93 | 7.981263 | 5274 |
| **2019-02-05 00:00:00** | 16.8 | 17.0651 | 16.8 | 16.8 | 7.919977 | 15532 |
| **2019-02-06 00:00:00** | 16.8 | 17.1899 | 16.8 | 17 | 8.014263 | 10645 |
| **2019-02-07 00:00:00** | 17.43 | 17.44 | 17.253 | 17.37 | 8.188691 | 11853 |
| **2019-02-08 00:00:00** | 17.65 | 17.88 | 17.65 | 17.831 | 8.406019 | 78599 |
| **2019-02-11 00:00:00** | 17.89 | 18.2199 | 17.89 | 18.13 | 8.546975 | 10287 |
| **2019-02-12 00:00:00** | 18.5465 | 19.02 | 18.5 | 18.53 | 8.735547 | 25820 |

Continued

| | | | | | | |
|---|---|---|---|---|---|---|
| **2019-02-13 00:00:00** | 18.56 | 18.74 | 17.96 | 18.4108 | 8.679354 | 70085 |
| **2019-02-14 00:00:00** | 18.49 | 18.52 | 18.3 | 18.4 | 8.674261 | 21213 |
| **2019-02-15 00:00:00** | 18.84 | 18.84 | 18.165 | 18.1862 | 8.57347 | 31359 |
| **2019-02-19 00:00:00** | 18.2 | 18.2 | 17.77 | 17.93 | 8.45269 | 14893 |
| **2019-02-20 00:00:00** | 17.79 | 18.5 | 17.79 | 18.1653 | 8.563616 | 23119 |
| **2019-02-21 00:00:00** | 18.37 | 18.37 | 17.9275 | 18.05 | 8.509261 | 14052 |
| **2019-02-22 00:00:00** | 18.33 | 18.355 | 18 | 18.3 | 8.627118 | 10887 |
| **2019-02-25 00:00:00** | 18.5 | 18.55 | 18.16 | 18.17 | 8.565834 | 21680 |
| **2019-02-26 00:00:00** | 18.1501 | 18.3899 | 17.81 | 18.3899 | 8.6695 | 16515 |

Table 2. Summary of the dataset.

| | Open | High | Low | Close | Adj Close | Volume |
|---|---|---|---|---|---|---|
| **Count** | 1258 | 1258 | 1258 | 1258 | 1258 | 1258 |
| **Mean** | 10.89 | 11.03 | 10.76 | 10.88 | 5.88 | 16892.58 |
| **Std** | 2.60 | 2.62 | 2.59 | 2.59 | 0.99 | 18906.56 |
| **Min** | 5.26 | 5.71 | 5.25 | 5.46 | 3.93 | 0 |
| **25%** | 8.91 | 9.03 | 8.84 | 8.93 | 5.16 | 6066.75 |
| **50%** | 10.13 | 10.28 | 10.01 | 10.14 | 5.73 | 11711.5 |
| **75%** | 12.48 | 12.62 | 12.31 | 12.45 | 6.56 | 21123.25 |
| **Max** | 18.84 | 19.02 | 18.5 | 18.53 | 8.74 | 202207 |

The dataset consists of 1258 data points for each column. The mean values for the stock's opening, highest, lowest, and closing prices are around 10.89, 11.03, 10.76, and 10.88, respectively. The mean adjusted close price is approximately 5.88, and the average trading volume is about 16892.58. The standard deviations indicate the degree of variation from the mean, with the highest variability observed in the trading volume.

## 3.3. Method and Procedure of Analysis

The analysis in this study involves comparing the performance of deep learning models (LSTMs, GRUs) with traditional statistical models (ARMA, ARIMA) in predicting stock prices in the Nigerian stock market.

### 3.3.1. Long Short-Term Memory (LSTM)

Long Short-Term Memory (LSTM) is a specialized form of recurrent neural network (RNN) designed to overcome the vanishing gradient problem, making it particularly effective at capturing long-term dependencies in sequential data such as stock prices. LSTMs are equipped with memory cells that can maintain information over long periods, enabling them to learn temporal patterns and relationships within the data. The LSTM architecture consists of several gates that regulate the flow of information:

Input Gate ( $i_t$ ): Determines how much of the new input to add to the cell state.

Forget Gate ( $f_t$ ): Decides what information to discard from the cell state.

Output Gate ( $o_t$ ): Controls the output and what information from the cell state to use for the hidden state.

The equations governing the behavior of an LSTM cell are as follows:

$$
\begin{aligned}
i_t &= \sigma\left(\mathcal{W}_{xi}x_t + \mathcal{W}_{hi}h_t + \mathcal{W}_{ci}c_{t-1} + 1\right) \\
f_t &= \sigma\left(\mathcal{W}_{xf}x_t + \mathcal{W}_{hf}h_t + \mathcal{W}_{cf}c_{t-1} - b_f\right) \\
g_t &= \tanh\left(\mathcal{W}_{xi}x_t + \mathcal{W}_{hg}h_{t-1} + b_g\right) \\
o_t &= \sigma\left(\mathcal{W}_{xo}x_t + \mathcal{W}_{ho}h_{t-1} + \mathcal{W}_{co}c_t + b_o\right) \\
c_t &= f_t \odot c_{t-1} + i_t \odot g_t \\
h_t &= o_t \odot \tanh\left(c_t\right)
\end{aligned}
\quad \text{(i)}
$$

where:

- $x_t$ represents the stock price input at time *t*.
- $h_t$ is the hidden state at time *t*.
- $c_t$ is the cell state at time *t*.
- $i_t$, $f_t$, $o_t$ and $g_t$ are the input, forget, output, and cell gates' activation vectors, respectively.
- $\sigma$ denotes the sigmoid function.
- $\odot$ denotes element-wise multiplication.

In the context of stock price prediction:

- **Input Gate** ( $i_t$ ): Determines how much of the current stock price and other relevant should influence the cell state.
- **Forget Gate** ( $f_t$ ): Decides how much of the previous cell state (previous stock prices and trends) should be retained.
- **Output Gate** ( $o_t$ ): Controls the amount of information from the cell state to be used in predicting the next stock price.

### 3.3.2. GRU (Gated Recurrent Unit)

Gated Recurrent Unit (GRU) is a simplified version of the Long Short-Term Memory (LSTM) network, designed to achieve similar results with fewer parameters, making it computationally more efficient. GRUs merge the memory and input gates into a single structure, streamlining the learning process while still capturing essential temporal dependencies in sequential data like stock prices.

The GRU architecture includes two main gates:

- Update Gate $z_t$: Controls how much of the past information needs to be passed along to the future.
- Reset Gate $r_t$: Determines how much of the past information to forget.

The equations governing the GRU cell are:

$$\begin{aligned} z_t &= \sigma\left(\mathcal{W}_{xz}x_t + \mathcal{W}_{hz}h_{t-1} + b_z\right) \\ r_t &= \sigma\left(\mathcal{W}_{xr}x_t + \mathcal{W}_{hr}h_{t-1} + b_r\right) \\ \tilde{h}_t &= \tanh\left(\mathcal{W}_{xh}x_t + f_t \odot \left(\mathcal{W}_{hh}h_{t-1}\right) + b_h\right) \\ h_t &= \left(1 - z_t\right) \odot h_{t-1} + z_t \odot \tilde{h}_t \end{aligned} \tag{ii}$$

where:

- $x_t$ represents the stock price input at time *t*.
- $h_t$ is the hidden state at time *t*.
- $z_t$ is the update gate's activation vector.
- $r_t$ is the reset gate's activation vector.
- $\tilde{h}_t$ is the candidate hidden state.
- $\sigma$ denotes the sigmoid function.
- $\odot$ denotes element-wise multiplication.

In the context of stock price prediction:

- Update Gate $z_t$: Controls how much of the previous stock price information should be retained.
- Reset Gate $r_t$: Determines how much of the previous information should be forgotten.
- Candidate Hidden State $\tilde{h}_t$: Incorporates new information and past data to predict future stock prices.

### 3.3.3. ARMA (Autoregressive Moving Average)

The Autoregressive Moving Average (ARMA) model is a powerful statistical tool used to predict future values in a time series by combining autoregressive (AR) and moving average (MA) components. This model is particularly effective in capturing the linear dependencies in sequential data, making it suitable for stock price prediction.

The ARMA model comprises two key components:

**Autoregressive (AR) Component:** This part of the model expresses the current value of the series as a linear combination of its previous values. The order of the AR component is denoted by *p*.

$$\text{AR}(p): y_t = \phi_1 y_{t-1} + \phi_2 y_{t-2} + \cdots + \phi_p y_{t-p} + \epsilon_t \tag{iii}$$

**Moving Average (MA) Component:** This part expresses the current value as a linear combination of past error terms. The order of the MA component is denoted by *q*.

$$\text{MA}(q): y_t = \epsilon_t - \theta_1\epsilon_{t-1} - \theta_2\epsilon_{t-2} - \cdots - \theta_p\epsilon_{t-q} \tag{iv}$$

Therefore, the combined ARMA model is represented as $\text{ARMA}(p,q)$:

$$y_t = \phi_1 y_{t-1} + \phi_2 y_{t-2} + \cdots + \phi_p y_{t-p} + \epsilon_t - \theta_1\epsilon_{t-1} - \theta_2\epsilon_{t-2} - \cdots - \theta_p\epsilon_{t-q} \tag{v}$$

where:

- $y_t$ represents the stock price at time t.
- $\phi$ and $\theta$ are coefficients to be estimated.
- $\epsilon_t$ is the white noise error term.

In the context of stock price prediction, the ARMA model utilizes past stock prices and error terms to forecast future prices. This approach helps in understanding the underlying patterns in stock prices by accounting for both the linear relationships (AR part) and the shocks or unexpected changes (MA part).

#### 3.3.4. ARIMA (Autoregressive Integrated Moving Average)

The AutoRegressive Integrated Moving Average (ARIMA) model is an extension of the ARMA model designed to handle non-stationary time series data by incorporating differencing. This makes ARIMA particularly effective for predicting stock prices, where trends and seasonal patterns often cause non-stationarity.

The ARIMA model has three main components.

AutoRegressive (AR) Component: Like in the ARMA model, the AR component uses previous values to predict the current value. The order of this component is denoted by *p*. So, from equation (iii):

$$\text{AR}(p): y_t = \phi_1 y_{t-1} + \phi_2 y_{t-2} + \cdots + \phi_p y_{t-p} + \epsilon_t$$

Integrated (I) Component: This part involves differencing the time series data to make it stationary. The order of differencing is denoted by *d*.

$$\text{I}(d): y'_t = y_t - y_{t-d} \quad \text{(vi)}$$

Moving Average (MA) Component: Similar to the ARMA model, the MA component uses past error terms to predict the current value. The order of this component is denoted by *q*. So, from equation (iv):

$$\text{MA}(q): y_t = \theta_1 \epsilon_{t-1} + \theta_2 \epsilon_{t-2} + \cdots + \theta_p \epsilon_{t-q} + \epsilon_t$$

Therefore, the combined ARMA model is represented as $\text{ARIMA}(p,d,q)$:

$$y'_t = \phi_1 y'_{t-1} + \phi_2 y'_{t-2} + \cdots + \phi_p y'_{t-p} + \theta_1 \epsilon_{t-1} + \theta_2 \epsilon_{t-2} + \cdots + \theta_p \epsilon_{t-q} + \epsilon_t \quad \text{(vii)}$$

where:

- $y_t$ represents the stock price at time *t*.
- $y'_t$ represent the difference stock price
- $\phi$ and $\theta$ are coefficients to be estimated.
- $\epsilon_t$ is the white noise error term.

In stock price prediction, the ARIMA model effectively captures both the linear relationships and the effects of trends and seasonality by differencing the data. This helps in stabilizing the time series and improving the accuracy of the forecasts.

### 3.4. Evaluation Metrics

To assess the accuracy and reliability of the models in predicting stock prices, this study employs several evaluation metrics. These metrics provide a quantitative basis for comparing the performance of deep learning models (LSTM, GRU) and traditional statistical models (ARMA, ARIMA). The metrics used include Mean Squared Error (MSE), Mean Absolute Percentage Error (MAPE), R-squared ($R^2$), and Root Mean Squared Error (RMSE).

### 3.4.1. Mean Squared Error (MSE)

Mean Squared Error (MSE) measures the average squared difference between the predicted and actual stock prices. It penalizes larger errors more significantly, making it sensitive to outliers. MSE is given by:

$$\text{MSE} = \frac{1}{n}\sum_{i=1}^{n}\left(P_i - A_i\right)^2 \quad \text{(viii)}$$

where:

- $n$ is the number of observations.
- $P_i$ is the predicted stock price at time $i$.
- $A_i$ is the actual stock price at time $i$.

In the context of stock prices, a lower MSE indicates that the model's predictions are closer to the actual prices, suggesting better performance.

### 3.4.2. Mean Absolute Percentage Error (MAPE)

Mean Absolute Percentage Error (MAPE) measures the average absolute percentage difference between the predicted and actual stock prices. It provides a relative measure of prediction accuracy, making it easy to interpret. MAPE is given by:

$$\text{MAPE} = \frac{1}{n}\sum_{i=1}^{n}\left|\frac{P_i - A_i}{A_i}\right| \times 100 \quad \text{(ix)}$$

MAPE is expressed as a percentage, with lower values indicating better predictive accuracy. It is particularly useful for comparing the performance of models across different datasets.

### 3.4.3. Root-Squared Error ($R^2$)

R-squared ($R^2$) is a statistical measure that represents the proportion of the variance in the dependent variable (actual stock prices) that is predictable from the independent variable (predicted stock prices). It indicates the goodness of fit of the model. $R^2$ is given by:

$$R^2 = 1 - \frac{\sum_{i=1}^{n}\left(A_i - P_i\right)^2}{\sum_{i=1}^{n}\left(A_i - \bar{A}_i\right)^2} \quad \text{(x)}$$

where:

- $\bar{A}_i$ is the mean of the actual stock prices.
- $P_i$ is the predicted stock price at time $i$.

An $R^2$ value closer to 1 indicates that the model explains a large portion of the variance in the stock prices, signifying a good fit.

### 3.4.4. Root Mean Squared Error (RMSE)

Root Mean Squared Error (RMSE) is the square root of the average squared difference between the predicted and actual stock prices. It is similar to MSE but is in the same units as the original data, making it more interpretable. RMSE is given by:

$$\text{RMSE} = \sqrt{\frac{1}{n}\sum_{i=1}^{n}\left(P_i - A_i\right)^2} \quad \text{(xi)}$$

Like MSE, a lower RMSE indicates better predictive performance. RMSE is particularly useful when the magnitude of the errors is important.

### 3.5. Analysis Approach

The analysis approach involves independently testing LSTM, GRU, ARMA, and ARIMA models on short-term (1 year), medium-term (2.5 years), and long-term (5 years) datasets. Figure 2 demonstrates the high-level abstraction of our analysis pattern. The Augmented Dickey-Fuller (ADF) test will be used to determine the stationarity of the data [21]. By assessing model performance across different time horizons, the study aims to provide insights into the models' effectiveness under varying conditions. This progressive evaluation facilitates a comprehensive comparison, shedding light on the predictive accuracy of deep learning and statistical models over diverse time frames.

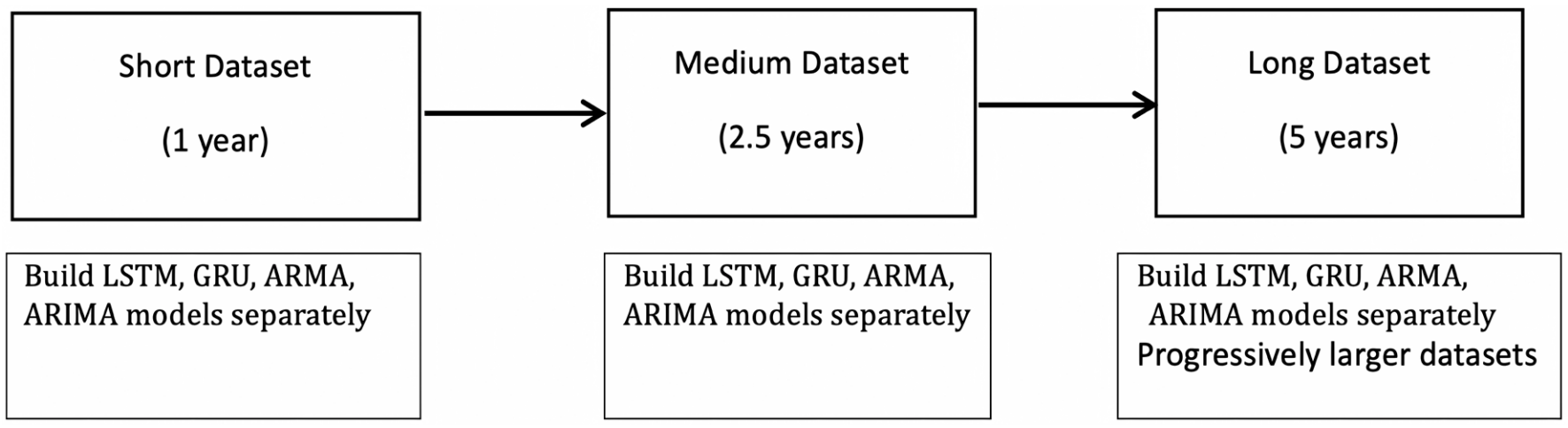

Figure 2. Analysis approach explanation.

## 4. Results

This analysis aims to evaluate the effectiveness of deep learning models, including RNNs like LSTMs and GRUs, compared to traditional ARMA and ARIMA models, for predicting stock prices on the Nigerian stock market. Additionally, the study assesses the influence of various input parameters, such as historical price data, trading volumes, economic indicators, and sentiment data, on the accuracy of stock price predictions.

Stock market charts are graphical representations of a stock or index's price movements over time. They are essential tools for investors and analysts to analyze trends, patterns, and potential opportunities in the market. Typically, the y-axis of a stock market chart represents the price of the stock or index, while the x-axis represents time.

The chart (Figure 3) depicting the Nigerian Exchange Stock Price over the past six years illustrates significant volatility in the market. It indicates that the stock price reached a peak of approximately 18 in 2021 and a low of around 6 in 2019, with the current trading price around 14. However, without further context or data, determining the specific reasons for these fluctuations is challenging. Possible factors influencing the stock price volatility could include changes in the Nigerian economy, fluctuations in interest rates, and global events impacting market sentiment. The chart's representation of 'Open' and 'Close' prices in Nigerian

Naira (NGN) suggests that it tracks the stock's daily price movements, providing viewers with insights into the stock's performance and potential trends. The Augmented Dickey-Fuller (ADF) test is commonly used to determine the stationarity of a time series dataset. The test statistic is compared to critical values at different confidence levels (1%, 5%, and 10%) to make this determination.

In this case, the ADF statistic is −35.73, which is significantly lower than the critical values at all confidence levels. Additionally, the p-value is reported as 0.0, indicating strong evidence against the null hypothesis of non-stationarity. Therefore, we reject the null hypothesis and conclude that the time series data for stock prices in the Nigerian stock market is stationary.

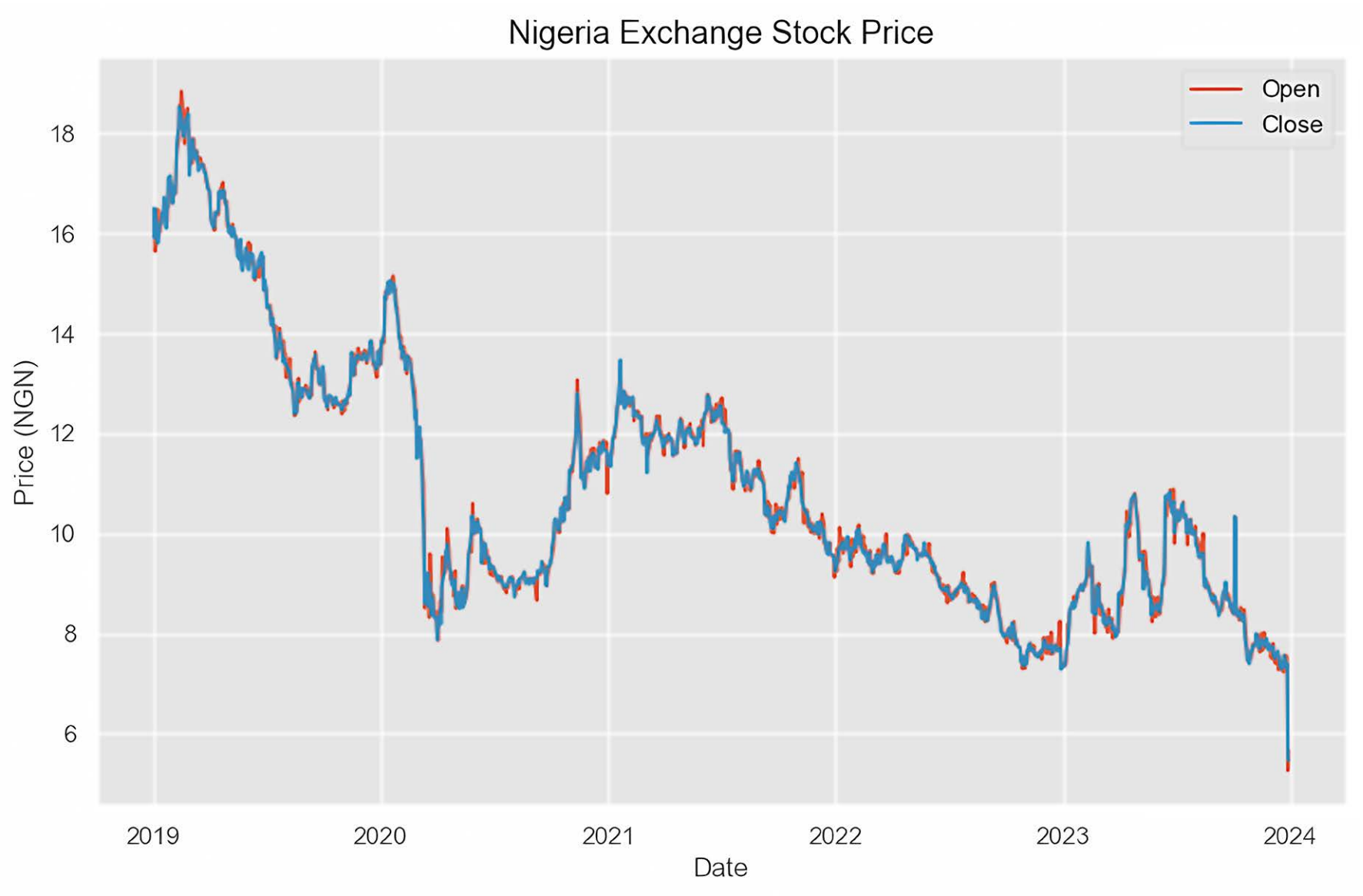

**Figure 3.** Nigeria stock exchange price (2019-2023).

The decomposition trend analysis presented in **Figure 4** provides insights into the price fluctuations of gold over time, likely denominated in Naira per ounce. The decomposition comprises three distinct sections: Original, Trend, and Residuals. The Original graph depicts the actual price of gold over an unspecified time frame, presumably spanning several years. The graph reveals notable peaks and troughs, indicative of price volatility and fluctuations over time.

The Trend line smooths out the inherent fluctuations in the original data, revealing a broader long-term trajectory. In this instance, the trend line demonstrates a modest upward slope, suggesting a gradual appreciation in the price of gold throughout the observed period.

Possible factors influencing the stock price volatility could include changes in the Nigerian economy, fluctuations in interest rates, and global events impacting market sentiment. The chart's representation of 'Open' and 'Close' prices in Nigerian Naira (NGN) suggests that it tracks the stock's daily price movements, providing viewers with insights into the stock's performance and potential trends.

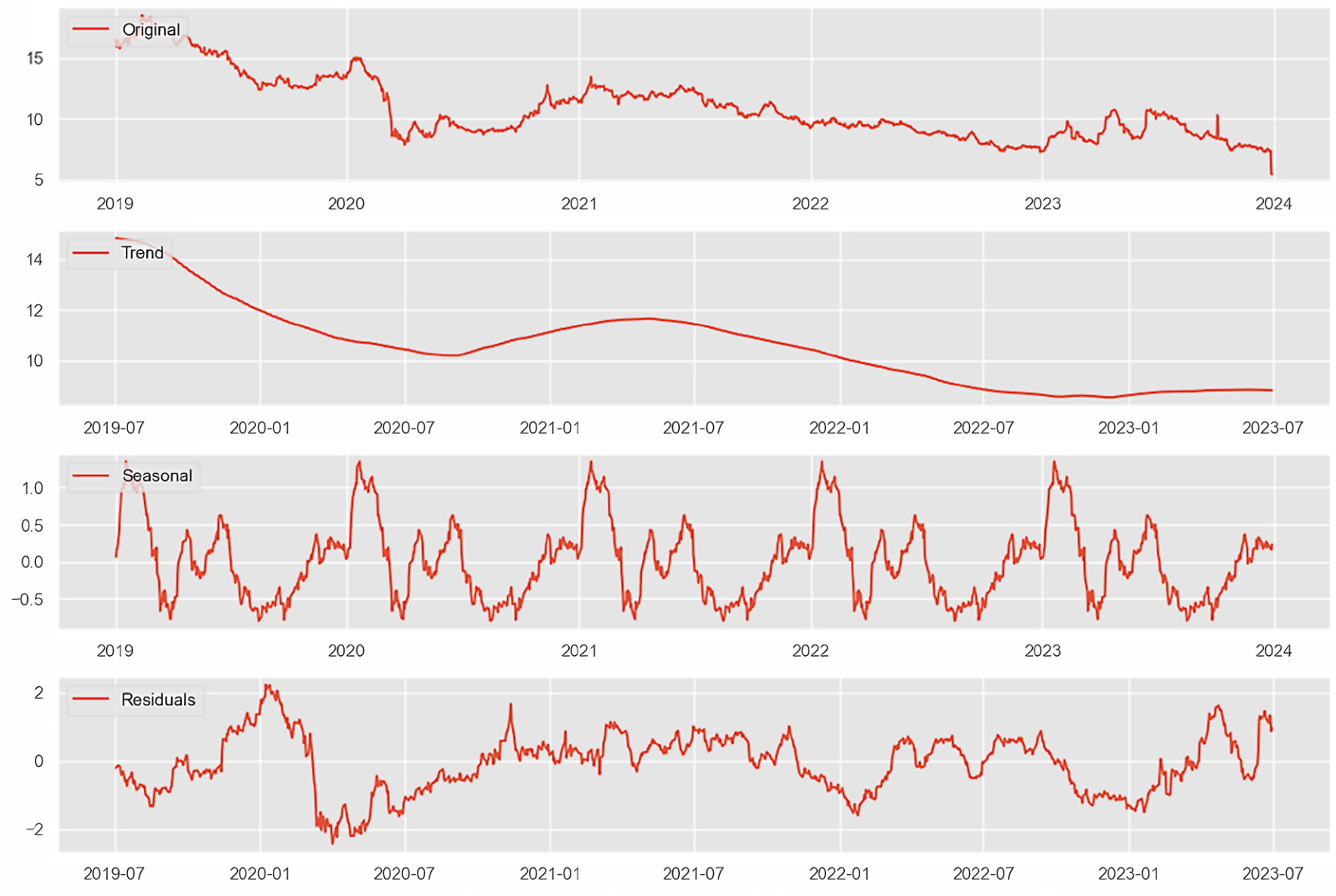

**Figure 4.** Decomposition trend.

## 4.1. Model Comparison

### 4.1.1. Short Term Analysis

In the short-term analysis of stock price prediction models, the performance of four models was evaluated: ARIMA, ARMA, LSTM, and GRU. The R-squared values indicate the goodness of fit of each model, with higher values indicating a stronger correlation between predicted and actual values. In Table 3**,** both ARIMA and ARMA models exhibited negative R-squared values, indicating poor performance compared to a simple mean model. This suggests that these traditional models were not effective in capturing the short-term movements of the stock price. On the other hand, the LSTM model showed impressive performance, with an R-squared value of 0.9885. This indicates a strong correlation between the predicted and actual values, suggesting that the LSTM model was highly effective in capturing short-term stock price movements. Similarly, the GRU model also performed well, with an R-squared value of 0.9796. This indicates that the GRU model was effective in capturing short-term stock price movements, although slightly less so than the LSTM model.

**Table 3.** Medium term analysis.

| Model | R-squared | RMSE | MSE | MAE |
|---|---|---|---|---|
| **ARIMA** | −1.7288 | 1.1707 | 1.3705 | 1.0171 |
| **ARMA** | −1.4705 | 1.1139 | 1.2408 | 0.9564 |
| **LSTM** | 0.9885 | 0.0761 | 0.0057 | 0.0569 |
| **GRU** | 0.9796 | 0.1013 | 0.0048 | 0.0970 |

In terms of error metrics, both the LSTM and GRU models had considerably lower Root Mean Squared Error (RMSE), Mean Squared Error (MSE), and Mean

Absolute Error (MAE) compared to the ARIMA and ARMA models. This further demonstrates the superior performance of the LSTM and GRU models in the short term.

Overall, the results suggest that the LSTM model was the best-performing model for short-term stock price prediction, followed closely by the GRU model. Moreover, Figure 5 shows the comparison of the predictions using different models that we discovered in short term analysis.

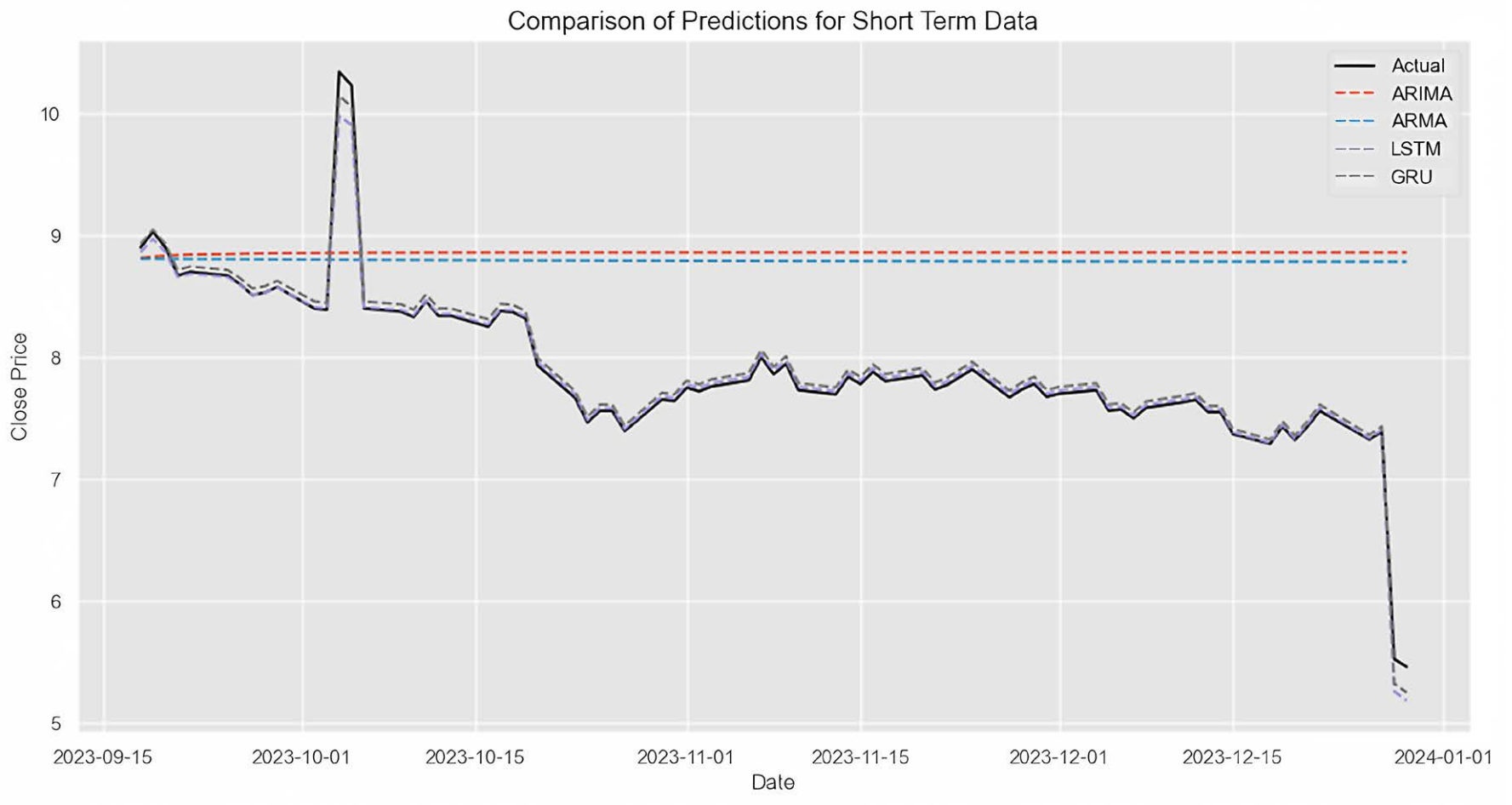

**Figure 5.** Predictions trend for short-term data.

### 4.1.2. Medium Term Analysis

In the medium-term analysis of stock price prediction models, the performance of ARIMA, ARMA, LSTM, and GRU models was evaluated. In Table 4, the ARIMA model showed a slight improvement in R-squared compared to the short term but still indicated poor performance. The R-squared value was −0.5725, suggesting that the ARIMA model was not effective in capturing medium-term stock price movements. The ARMA model performed significantly worse in the medium term, with a very low R-squared value of −10.9650. This indicates that the ARMA model is not suitable for medium-term stock price prediction.

In contrast, both the LSTM and GRU models performed well in the medium term. The LSTM model had an R-squared value of 0.8665, indicating a strong correlation between predicted and actual values. The GRU model performed slightly worse but still effectively captured medium-term stock price movements, with an R-squared value of 0.1256.

**Table 4.** Medium term analysis.

| Model | R-squared | RMSE | MSE | MAE |
|---|---|---|---|---|
| ARIMA | −0.5725 | 0.4018 | 0.1615 | 0.2961 |
| ARMA | −10.9650 | 1.1084 | 1.2285 | 0.9567 |
| LSTM | 0.8665 | 0.1171 | 0.0137 | 0.1066 |
| GRU | 0.1256 | 0.2996 | 0.0898 | 0.2959 |

In terms of error metrics, both the LSTM and GRU models had lower RMSE, MSE, and MAE compared to the ARIMA and ARMA models, further confirming their effectiveness in the medium term. Overall, the results suggest that the LSTM model was the best-performing model for medium-term stock price prediction, followed by the GRU model. Furthermore, Figure 6 depicts the comparison of the predictions using different models that we discovered in medium term analysis.

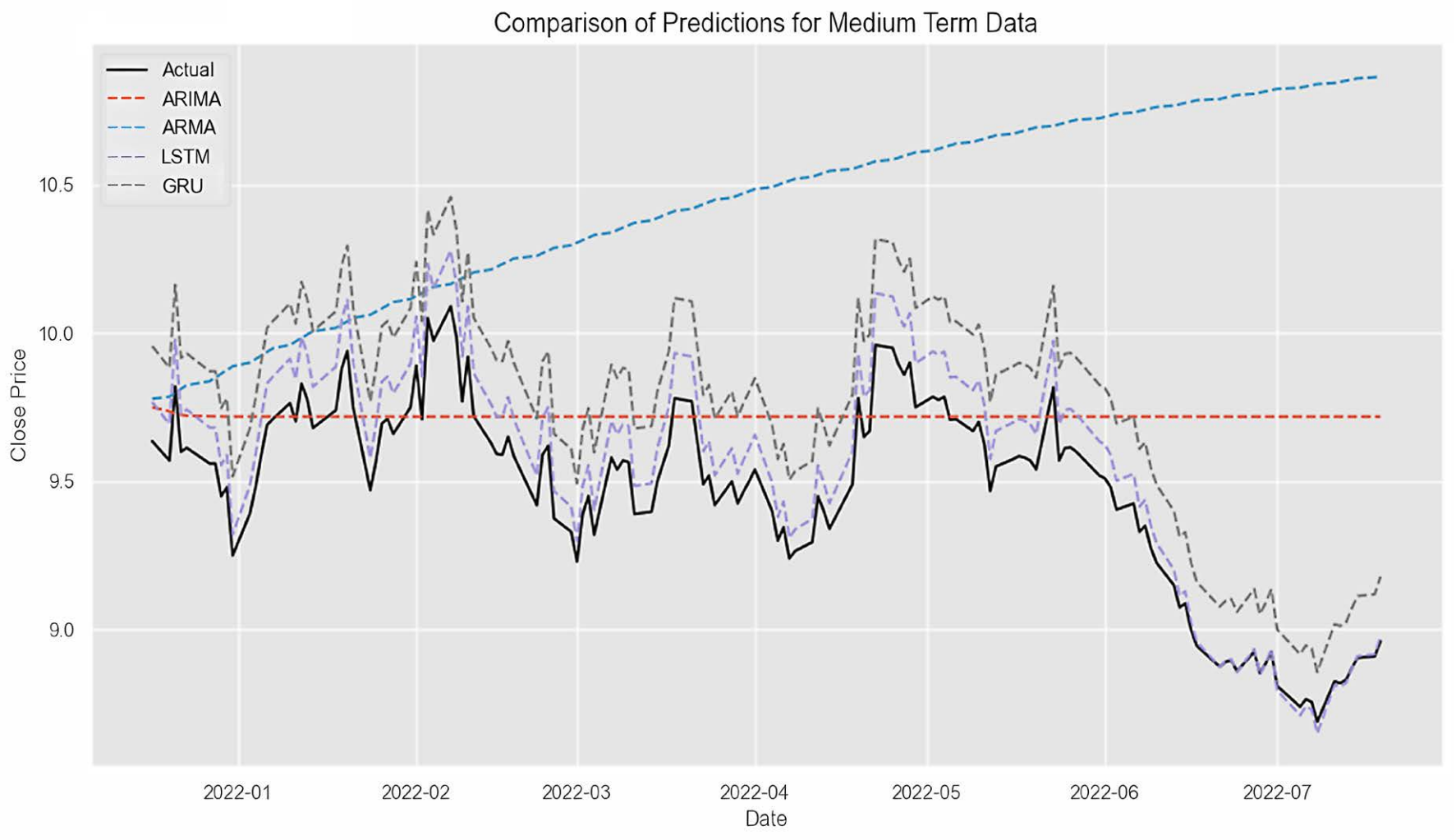

Figure 6. Predictions trend for medium-term data.

#### 4.1.3. Long Term Analysis

In the long-term analysis of stock price prediction models, the performance of ARIMA, ARMA, LSTM, and GRU models was evaluated. In Table 5, the ARIMA model exhibited an R-squared value of −2.3549, indicating poor performance in capturing long-term stock price movements. Similarly, the ARMA model performed poorly in the long term, with an R-squared value of −7.4547. In contrast, both the LSTM and GRU models showed much better performance in the long term. The LSTM model had an R-squared value of 0.4950, indicating a moderate correlation between predicted and actual values over the long term. The GRU model performed even better, with an impressive R-squared value of 0.9594, indicating a strong correlation between predicted and actual values in the long term. In terms of error metrics, both the LSTM and GRU models had lower RMSE, MSE, and MAE compared to the ARIMA and ARMA models, further confirming their effectiveness in the long term. Also, Figure 7 depicts the comparison of the predictions using different models that we discovered in long term analysis.

Table 5. Long term analysis.

| Model | R-squared | RMSE | MSE | MAE |
|---|---|---|---|---|
| ARIMA | −2.3549 | 0.9607 | 0.9229 | 0.8219 |
| ARMA | −7.4547 | 1.5250 | 2.3257 | 1.3248 |
| LSTM | 0.4950 | 0.3727 | 0.1389 | 0.3703 |
| GRU | 0.9594 | 0.1056 | 0.0111 | 0.0922 |

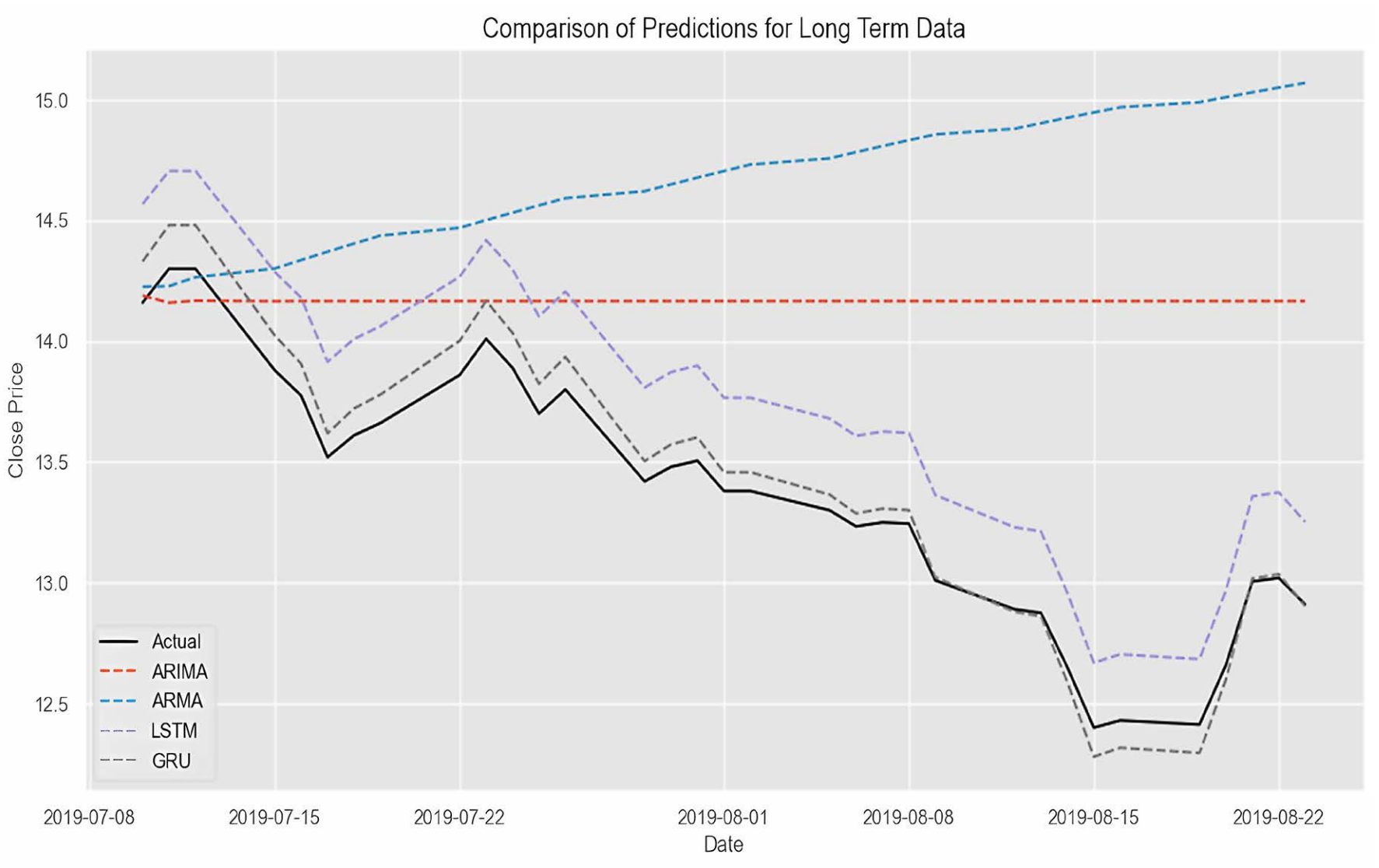

**Figure 7.** Predictions trend for long-term data.

In summary, the results suggest that the GRU model was the best-performing model for long-term stock price prediction, followed by the LSTM model.

## 5. Discussion

In order to anticipate stock prices on the Nigerian stock market, a comparative comparison of deep learning models (LSTM, GRU) and conventional statistical methods (ARIMA, ARMA) offers important new insights into the predictive modeling potential of machine learning and generative AI in the future. The superior performance of deep learning models especially LSTM over the traditional techniques is consistent with their capacity to extract intricate, nonlinear correlations from financial data. This implies that the use of machine learning (ML) techniques can greatly improve prediction accuracy across a range of domains. This is corroborated by recent research, which shows how ML models like ANN, SVM, and fuzzy logic are successfully applied in fields like crop production prediction, underscoring the technology' wider applicability [22] [23].

Furthermore, a new frontier for predicting and simulation is presented by the integration of generative AI (GenAI), which is quickly gaining popularity in both academic and industry contexts. Finance, agriculture, and climate science are just a few of the sectors that stand to gain from using GenAI's capacity to create intricate scenarios based on past data to create predictive models that are more resilient and adaptive [23]. The ability of GenAI to generate precise simulations on its own has enormous potential since it enables data-driven, real-time decision-making in a variety of disciplines that can maximize results and handle uncertainty. Predictions' future, then, will come from combining conventional machine learning techniques with the creative powers of artificial intelligence (GenAI), which will provide more accurate, scalable, and context-aware answers to challenging problems [23] [24]. We will explore that in our future research.

## 6. Conclusions

The research undertaken in this study focused on comparing the effectiveness of deep learning models and traditional models in predicting stock prices. The primary objective was to evaluate their performance, identify strengths and weaknesses, and understand their practical applications in financial markets.

The literature review revealed that while traditional models such as ARIMA and ARMA have been widely used, they face limitations in capturing the complex and nonlinear relationships in stock price data, particularly in volatile market conditions. On the other hand, deep learning models, including neural networks and LSTM, have shown promise in capturing these intricate patterns and dynamics. The methodology employed in this study involved data preprocessing, model implementation, and evaluation using metrics such as Mean Squared Error (MSE) and Mean Absolute Error (MAE). The Augmented Dickey-Fuller (ADF) test was used to assess the stationarity of the stock price series, which is crucial for effective modeling.

The analysis and results demonstrated that deep learning models generally outperformed traditional models in predicting stock prices across short-, medium-, and long-term horizons. The superior performance of deep learning models can be attributed to their ability to capture complex patterns and nonlinear relationships in the data. However, it was also noted that deep learning models require substantial computational resources and are less interpretable compared to traditional models. The findings of this study have significant implications for various stakeholders in the financial industry. For investors and financial analysts, the insights gained from this research can aid in making more informed investment decisions by leveraging the predictive power of deep learning models. However, it is essential to understand the limitations and potential risks associated with relying solely on these models. From a policy perspective, the study highlights the need for regulatory frameworks that encourage innovation in financial technologies while mitigating potential risks. Policymakers can use these insights to develop guidelines and standards for the responsible use of predictive models in the financial industry. The study also acknowledges certain limitations, such as the reliance on historical data, which may not always reflect future market conditions. Additionally, the computational intensity of deep learning models may limit their practical application for some users.

By offering a comparison of deep learning and conventional models for stock price prediction, this research advances financial theory and practice overall. However, acknowledging the utility of traditional models in some situations, it also highlights the promise of deep learning in capturing intricate market dynamics. In the end, the results promote better decision-making and increase market efficiency by providing investors, analysts, and regulators with actionable insights.

## Conflicts of Interest

The authors declare no conflicts of interest regarding the publication of this paper.